\documentclass[twocolumn,showpacs,preprintnumbers,amsmath,amssymb]{revtex4}
\usepackage{graphicx}
\input epsf
\usepackage{dcolumn}
\usepackage{bm}

\begin{document}


\title{The dressing procedure for the cosmological  equations and the indefinite future of the universe}
\author{A.V. Yurov}
\email{artyom_yurov@mail.ru}
{%
\affiliation{%
I. Kant Russian State University, Theoretical Physics Department,
 Al.Nevsky St. 14, Kaliningrad 236041, Russia
\\
 }

\author{A.V. Astashenok}
\email{artyom.art@gmail.com}
{%
\affiliation{%
I. Kant Russian State University, Theoretical Physics Department,
 Al.Nevsky St. 14, Kaliningrad 236041, Russia;
\\
 }
\author{V.A. Yurov}%
 \email{valerian@math.missouri.edu}
 \affiliation{%
Department of Mathematics, University of Missouri, Columbia, MO
65211, U.S.A.
\\
}%


\date{\today}


\begin{abstract}
In this paper we present a new simple method of construction of infinite number of solutions
of Freidmann equations from the already known ones, which allows for a startling conclusion
of practical impossibility of correct predictions on the universe's future dynamics which
are based solely on astronomical observations on the value of a scale factor. In addition,
we present particular examples of newly constructed solutions, such as the ones, describing
the smooth dynamical (de)-phantomization,
and the models lacking the events horizons (both in classical and brane world cases).
The generalization of the method to the simplest anisotropic universes are presented as well.
\end{abstract}


\maketitle

\section{Introduction}

Imagine for a second an immortal astronomer, restlessly surveying an
expansion of the universe, how it goes during $10^{20}$ consequent
years. Suppose that she already knows for sure about the
flatness of her universe. Then, as time goes, the astronomer will
gather an (impressive) amount of observational data, visually
presented as a set of points on the graph $\left\{t_i,
a(t_i)\right\}$. If those points are located with appropriate
density, then it should be possible to choose some continuous
function $a(t)$, which with high precision will approximate the
gained set. Of course, this function must not be unique, but let’s
assume for now that our astronomer have managed (not without some
efforts) to find out the most satisfactory one; when this is done
she might conclude that now she posses the actual time dependence
of observed scale factor. Moreover, since both matter’s density
and pressure are expressed in terms of scale factor (and it’s
derivatives) explicitly (we remind here, that under the assumption
the sign of curvature is already known):
\begin{equation}
\rho=\frac{3{\dot a}^2}{8\pi Ga^2},\qquad
p=-\frac{c^2\left(2{\ddot a}a+{\dot a}^2\right)}{8\pi Ga^2},
\label{oh}
\end{equation}
 then she might as well
conclude that the observations have given her the values of all basic
characteristics of the universe. Can we say now that the further
evolution of her universe will be completely defined? No, we can't.

This paper contains the discussion of an extremely simple method which allows
one to construct new solutions of the cosmological Friedmann
equations starting out from the previously known (and more simple) ones.
The method itself will be denoted as the method of linearization, and the
transformations which allows one to construct new solutions
starting out from another ones will be refereed to as
''the dressing'' or as a ''dressing procedure''. What is new and interesting
about such transformations is that they admit the {\it
invariants}. More precisely, the special linear combination of the
density $\rho$ and pressure $p$ (of a form $U=\alpha\rho+\beta p$) is
covariant with respect to these transformations ($\alpha\rho+\beta
p=\alpha{\tilde \rho}+\beta {\tilde p}$). (For example, in one particular case
${\tilde p}=p$).

There are in total four aims of this paper. First of all, we will
show that for any function $a(t)$ there exist such solutions
$a_n$, which, being similar to $a(t)$ at given time interval with
any given precision rate, will completely differ from $a(t)$ later
(note here, that possibility of such scenario, i.e. of
indeterminacy of universe future has been previously noted by A.
A. Starobinsky in \cite{Starob}). These solutions will be
constructed via the dressing procedure starting out from the
initial solution $a(t)$ (Sec. II). Secondly, as we shall see in
Section III, upon further usage of a dressing procedure one can
construct special solutions, describing the smooth dynamical
(de)-phantomization. Thirdly, in Sec. IV we will show that
described transformations allow one to construct space-times
without event horizons. Finally, in Sec. V we will take a brief look on possible generalizations of presented method
for the simplest anisotropic cosmological model and the particular
Randal-Sundrum-I brane model and will show that in both cases the similar dressing procedures would still be 
valid and they yield the same results as the ones described above.

\section{Dressing procedure}

Let us write the Friedmann equations as
\begin{equation}
\left(\frac{\dot a}{a}\right)^2=\rho-\frac{k}{a^2},\qquad
2\frac{{\ddot a}}{a}=-(\rho+3p), \label{Fried}
\end{equation}
where $k=0,\pm 1$. Throughout the paper (excluding Sec.V) we'll stick to the
metric units with $8\pi G/3=c=1$.
If the universe is filled with a self-acting  and minimally
coupled scalar field with Lagrangian
\begin{equation}
L=\frac{{\dot\phi}^2}{2}-V(\phi)=K-V, \label{Lagrangian}
\end{equation}
then the energy density and pressure are
$$
\rho=K+V,\qquad p=K-V,
$$
therefore
\begin{equation}
V=\frac{1}{2}(\rho-p),\qquad K=\frac{1}{2}(\rho+p). \label{Vphi}
\end{equation}
Our starting point is that the volume function $\psi=a^3$
satisfies a simple second-order differential equation
(\cite{1Chervon})
\begin{equation}
\ddot\psi = 9V\psi. \label{privet}
\end{equation}
In (\ref{privet}) the potential $V$ is represented as a function
of time $t$. For simple forms of the potential one can find the
general solution of (\ref{privet}), containing both the solution used for the construction of this potential and
a lineary independent one as well. Substituting this general solution
into the (\ref{Fried}) one can calculate $\rho$ and $p$. Then
using (\ref{Vphi}) one gets the new potential $\tilde V$ such that
${\tilde V}(t)=V(t)$  but whose form is different from $V$ if
$\tilde V$ and $V$ are represented as functions of $\phi$:
${\tilde V}({\tilde\phi})\ne V(\phi)$. This simple method allows
one to construct cosmological models describing a smooth
transition from ordinary dark energy to the phantom one
(\cite{YV}, \cite{Andr}).

Therefore the Friedmann  equations admits the linearizing
substitution and can be studied via  different powerful
mathematical methods which were developed for the linear
differential equations. This is the reason why we call our
approach the method of linearization. The crucial point of this
paper is connected to the simple generalization of results above.
More precisely, the following proposition is hold:
\newline
{\bf Proposition.} Let $a=a(t)$ (with $p=p(t)$, $\rho=\rho(t)$) be
the solution of (\ref{Fried}) with  $k=0$. Then the function
$\psi_n\equiv a^n$ is the solution of the Schr\"odinger equation
\begin{equation}
\frac{{\ddot \psi_n}}{\psi_n}=U_n, \label{Scr}
\end{equation}
with potential
\begin{equation}
U_n=n^2\rho-\frac{3n}{2}\left(\rho+p\right). \label{Un}
\end{equation}
For example:
$$
U_1=-\frac{\rho+3p}{2},\qquad U_2=\rho-3p,\qquad
U_3=\frac{9}{2}(\rho-p),
$$
or
$$
U_{1/2}=-\frac{1}{2}\left(\rho+\frac{3p}{2}\right),\qquad
U_{-1}=\frac{5\rho+3p}{2}
$$
and so on.
\newline
{\it Remark 1}. If the universe is filled with scalar field $\phi$
whose Lagrangian is (\ref{Lagrangian}) then the expression
(\ref{Un}) will be
$$
U_n=n(n-3)K+n^2V.
$$
In particular case $n=3$ $U_3=9V(\phi)$ (see (\ref{privet})). This
particular case has been extensively studied in \cite{1Chervon}, \cite{YV}.
\newline
{\it Remark 2}. For small values of $n\ll 1$ one gets $U_n\sim
-3n(\rho+p)/2$; for example, if $n=0.01$ then
$$
U_n\sim -(0.0149\rho+0.015p)\sim -0.015(\rho+p).
$$
Therefore one can use $U_n<0$ to check whether the weak energy
condition is violated~\footnote{At the same time, one shall keep in
mind that this eqaution is nothing but approximate. To ensure (with the help of $\psi_n$) that the
weak energy condition will indeed be violated, one can
use exact equation ${\dot\sigma_n}=-3n(\rho+p)/2$, where
$\sigma_n={\dot\psi_n}/\psi_n$.}. If, on the contrary, $n\gg 1$
then $U_n\sim n^2\rho$.

In the case of general position, the solution of the equation
(\ref{Scr}) has the form
\begin{equation}
\Psi_n=c_1\psi_n+c_2{\hat\psi}_n, \label{total}
\end{equation}
where ${\hat\psi}_n$ is linearly independent counterpart of
$\psi_n$:
\begin{equation}
{\hat\psi}_n(t)=\psi_n(t)\int^t \frac{dt'}{\psi_n^2(t')}\equiv
\psi_n(t)\xi(t). \label{xi}
\end{equation}
Equation (\ref{total}) is enough to establish the following
theorem:
\newline
{\bf Linearization Theorem}. Let $a=a(t)$ be the solution of (\ref{Fried}) with
$k=0$ and with $\rho$ and $p$, given by (\ref{oh}). Then the two-parameter function
$a_n=a_n(t;c_1,c_2)$:
\begin{equation}
a_n=a\left(c_1+c_2\int\frac{dt}{a^{2n}}\right)^{1/n}, \label{an}
\end{equation}
will be solution of  (\ref{Fried}) with new energy density
$\rho_n$ and pressure $p_n$ satisfying:
\begin{equation}
n^2\rho_n-\frac{3n}{2}\left(\rho_n+p_n\right)=
n^2\rho-\frac{3n}{2}\left(\rho+p\right). \label{inv}
\end{equation}
Another way to formulate this theorem is to say that the
expression
$$
U_n=\frac{n\left((n-1){\dot a}^2+a{\ddot a}\right)}{a^2},
$$
is invariant with respect to transformation $a\to a_n$ with $a_n$
defined by (\ref{an}). We'll use the term ''dressing'' for the
process of transformation of a triple $\{a,\rho,p\}$, with the
resulting triple $\{a_n,\rho_n,p_n\}$ being referred to as the
dressed  one.
\newline
{\it Remark 3}. This theorem is valid for the case $k=0$. If
$k=\pm 1$ then this theorem will hold if and only if  $n=0,1$.
\newline
{\it Remark 4}. If $n=3/2$ then the equation (\ref{inv}) results
in $p_n=p$ (but $\rho_n\ne\rho$).

Let $a(0)=0$, i.e. suppose that  at $t=0$ there exist an initial singularity. Lets
assume that $a(t)\sim t^{\lambda}$ for $t\to 0$. One might easily
verify that if $2n\lambda \leq 1$ then $a_n(0)=0$. Now let us choose
$|c1/c_2|\gg |\xi_{\rm max}|$ where $\xi=\xi(t)$ is the quantity from
(\ref{xi}); $\xi$ is a bounded function at the interval $0<t<T$ and
$\xi_{\rm max}$ is the maximal value of $\xi(t)$ at this
interval.  It can been seen that at a given time interval
$a_n(t)$ behaves similar to $a(t)$ with any given precision rate.

Therefore, the conjecture of our immortal astronomer was wrong: Her observations
might result in $a(t)$, but she can't be sure that the ''real'' scale
factor is $a(t)$. It can be $a_n(t)$ as well.

\section{Smooth dynamical (de)-phantomization}

Let us now impose the equation of state:
$p=w\rho=(\gamma-1)\rho$, where $\gamma$ is adiabatic index. In
this case the solution of equations (\ref{Fried}) with $k=0$ has
the form
$$
a=a_it^{2/(3\gamma)},\qquad \rho=\frac{4}{9\gamma^2t^2},
$$
where $a_i=$const. Using ''dressing procedure'' $a\to a_n$ one gets
a new solution $a_n=a_n(t;c_1,c_2)$ such that
$$
a_n(t;c_1,0)=c_1^{1/n}a(t),\qquad a_n(t;0,c_2)={\tilde
a_i}t^{2/(3{\tilde \gamma})},
$$
where
$$
{\tilde
a_i}=\frac{1}{a_i}\left(\frac{3c_2\gamma}{3\gamma-4n}\right)^{1/n},\qquad
{\tilde \gamma}=\frac{2\gamma n}{3\gamma-2n},
$$
or
$$
{\tilde w}=\frac{2nw+4n-3(w+1)}{3(w+1)-2n}.
$$
Phantom energy always takes place for negative $\gamma$ (or $\tilde\gamma$).
If, on the other hand, $\gamma>0$ then $a_n(t;0,c_2)$ will describe the
phantom cosmological model whenever $n$ lies in the range $n<0$ or
$n>3\gamma/2$. For example, if initial $a(t)$ describes the dust
universe with $w=0$ ($\gamma=1$) then $a_n(t;0,c_2)$ will
describe the phantom cosmological model for $n<0$ and $n>3/2$; if
the initial scale factor describes radiation universe with $w=1/3$
then one gets the phantom model both for negative $n$ and $n>2$.

If $c_1\ne 0$ and $c_2\ne 0$ then (\ref{an}) will describe the
smooth dynamical (de)-phantomization (see \cite{YV} for the case
$n=3$). Phantom energy results in big rip singularity at some time
$t=t_*$. Controlling the zeroes of the starting solutions $a(t)$,
we can construct exact cosmologies with the presence of big rip. In fact, let
$a(t_*)=0$ and $a(t)\sim (t-t_*)^{\lambda}$ or $a(t)\sim
(t_*-t)^{\lambda}$ for $t\sim t_*$. Using (\ref{an}) one can
conclude that, for $n\lambda>1$, $a_n(t;c_1,c_2)$ will describe the
big rip as the singularuty of $a(t)$ is approached.

If $\xi(t)\to 0$ as $t\to\pm\infty$ then one has a solution
allowing for finite ''phantom regions''. In other words, there exist
two instants $t_1$ and $t_2$,   $t_1<t_*<t_2$  such that
$(\rho_n+p_n)(t_1)= (\rho_n+p_n)(t_2)=0$ and $\rho_n+p_n<0$ if
$t_1<t<t_2$.

All these regions can be cut out, with the edges being newly sewn
up in such a way that neither the scale factor (and its first two
derivatives) nor density or pressure will experience a jump anymore. To show
this let us assume that the universe is born at some time instant $t=0<t_1$
and suppose there exist a time interval, during which all the
energy conditions are satisfied. One can demonstrate that as time
goes to $t_1$, the strong energy condition is the one bounded to be violated
first, the weak one being extant. As a result, the universe would
get inflated up to the instant $t_1$. Furthermore, when $t=t_1$ we
arrive precisely to the De Sitter's phase, but this time {\em
instantaneously}. What follows next is the "phantom region",  {\bf
which can be cut out} by matching the solutions corresponding to
$t=t_1$ and $t=t_2$.  Note that gluing the solutions in such a fasion
we shall make sure to use the constraints much stricter than those
usually imposed for the similar problems of non-relativistic quantum
theory based on the Schr\"odinger equation. In fact, what the discussed problem
calls for is a $C^2$ gluing; namely the functions has to be matched
up to their second derivatives. The latter requirement serves to
eliminate the possibility of discontinuities of pressure and
density. A general solution of equation (\ref{Scr}) is
characterized by two integration constants, and it is not obvious
that three matching conditions will be satisfied in general. However, it
is easy to verify that the third condition, namely the equality of
the second derivatives, is {\bf automatically fulfilled}, provided
that the first two, namely the $C^1$ matching has been successfully applied. This
apparent fluke is due to the fact that the matching is done at an
inflection point of $\log\psi$.

Indeed, let us return to (\ref{total}). It is possible to write
$\Psi_n^{(\pm)}=c^{(\pm)}_1\psi_n+ c^{(\pm)}_2{\hat \psi}_n$ where
$\Psi_n=\Psi_n^{(-)}$ at $t<t_1$ and
$\Psi_n=\Psi_n^{(+)}$ at $t>t_2$. Using the conditions
$$
\Psi_n^{(-)}(t_1)=\Psi_n^{(+)}(t_2),\qquad {\dot
\Psi}_n^{(-)}(t_1)={\dot \Psi}_n^{(+)}(t_2),
$$
one can express the constants $c^{(+)}_{1,2}$ via
$c^{(-)}_{1,2}$, $a(t_{1,2})$ and ${\dot a}(t_{1,2})$. Since
$\rho_n=n^2{\dot a}^2/a^2$, it follows that $\rho_n(t_1)=\rho_n(t_2)$. But
$p_n(t_{1,2})=-\rho_n(t_{1,2})$, therefore $p _n(t_1)=p _n(t_2)$.
Using (\ref{Scr}) one concludes that $a_n(t_1)=a_n(t_2)$, ${\dot
a}_n(t_1)= {\dot a}_n(t_2)$ and ${\ddot a}_n(t_1)= {\ddot
a}_n(t_2)$, and hence
${\dot\rho}(t_1)={\dot\rho}(t_2)$~\footnote{But this would not be true for pressure: ${\dot p}_n(t_1)\ne{\dot p}_n(t_2)$!}.

Therefore, the finite phantom region can be cut out. However, an immediate question
arises here: is it really \textit{required} to do this? Of course, the smooth
phantomization looks extremely suspicious. In the universe filled with
scalar field $\phi_n$ ($\phi_n$ is the dressed field:
$\phi\to\phi_n$) it leads to the highly unusual situation: ${\dot\phi}_n^2>0$ at
$t<t_1$, $t>t_2$ while ${\dot\phi}_n^2<0$ at $<t_1<t<t_2$. Is it possible for
kinetic term to change its sign? While absolutely impossible in Minkowski
space, this may still be the case in the General Relativity
\cite{Andr} (see also \cite{Andr2},\cite{Odin}). Our point of view is this: one
shall treat the smooth phantomization as one real physical
phenomenon. If, however, the finite phantom region is an undesirable one then it can
simply be cut out by the method described above.

\section{Space-time without events horizon}

The equation (\ref{an}) allows one to construct the space-time
without events horizon (see also \cite{Tipler}). In order to show this let $t_f$ be an instant, such that
$a_n(t_f;c_1,c_2)=0$ but $a(t_f)=a_f\ne 0$. In other words, suppose that
$c_1+c_2\xi_f=0$, where $\xi_f=\xi(t_f)$. Then, for $t\to t_f$
(and $t<t_f$) one gets
\begin{equation}
a_n\sim a_f\left(c_1+c_2\xi_f-c_2(t_f-t){\dot\xi}_f\right)^{1/n}=
\frac{\kappa}{a_f}\left(t_f-t\right)^{1/n}, \label{OPT}
\end{equation}
where $c_2=-\kappa^2<0$. Integrating this equation for future
directed radial null geodesics, $ds^2=dt^2-a_n^2d\chi^2$ one will get
$$
\Delta\chi\sim\frac{a_f}{\kappa}\int^{t_f}\frac{dt'}{(t_f-t)^{1/n}}\sim
\frac{na_f}{\kappa(1-n)} \lim_{_{t\to t_f}}(t_f-t)^{(n-1)/n}.
$$
It is easy to see that for $0<n\le 1$ we will have $\Delta\chi=+\infty$
which shows that radial null geodesics circumnavigate the universe
infinite number of times as the future c-boundary at $t = t_f$
is approached. By homogeneity and isotropy, we can conclude that
all future endless timelike curves define the same c-boundary
point.

In the case $n=1$ one gets $\Delta\chi\sim
-\log\left(t_f-t\right)\to+\infty$ as the final singularity is
approached. This case is extremely interesting because (see Remark
3) when $n=1$ one can use the dressing procedure to construct the exact
solutions for the universes with $k=\pm 1$. Let's consider a few
examples.

\subsection{\label{sec:level2}Simple generalization of $k=+1$ dust model}

In the simplest dust case with $p=0$ one can solve the system
(\ref{Fried}) to get
\begin{equation}
\begin{array}{l}
a=a_m\sin^2\eta,\qquad
\displaystyle{2\eta-\sin 2\eta=\frac{2t}{a_m}},\\
\\
\displaystyle{\rho=\frac{1}{a_m^2\sin^6\eta}}.
\end{array}
\label{sol-1}
\end{equation}

Using (\ref{an}) for the case $n=1$ one will obtain the general
solution
\begin{equation}
\displaystyle{ a(t)_{gen}=c_1a(t)+c_2{\hat a}(t)}, \label{gen-1}
\end{equation}
where $c_{1,2}$ are the arbitrary constants. It is possible to rewrite
(\ref{gen-1}) in the form
\begin{equation}
a(t)\equiv a(t)_{gen}=A\sin\eta\sin\left(\delta-\eta\right),
\label{obs}
\end{equation}
with two arbitrary constants $A$ and $\delta$. This solution
describes the universe being born from the initial singularity
($\eta_i=0$, $t_i=0$) and collapsing thereafter into the final singularity at
$\eta_f=\delta$ or
$$
t_f=\frac{1}{2\alpha}\left(2\delta-\sin 2\delta\right)
$$
where we have introduced a new parameter $\alpha$ such that
$$
2\eta-\sin 2\eta=4\alpha t,\qquad
\alpha=\frac{1}{2a_m}\sin^2\frac{\delta}{2};
$$
hence,
\begin{equation}
\displaystyle{a=\frac{a_m}{\sin^2(\delta/2)}\sin\eta\sin\left(\delta-\eta\right),}
\label{sol}
\end{equation}
and the maximum value of $a=a_m$ will occur at $\eta=\delta/2$.

 It is easy to see that upon the choice $\delta=\pi$ one gets the well known
''dust solution'' (\ref{sol-1}). In case of general position
one shall choose $0<\delta<\pi$. It can be seen that for $t\ll t_f$,
$\delta=\pi-\epsilon$ and $\epsilon\ll 1$, (\ref{sol}) will behave
similar to (\ref{sol-1}). But if $t\sim t_f$ then one gets something
really different: a universe without events horizon. To show this
lets consider
$$
ds^2=dt^2-\frac{1}{4\alpha^2}\sin^2\eta\sin^2(\delta-\eta)\left[d\chi^2+\sin^2\chi
d\Omega^2_3\right].
$$
Upon integration of the equation $ds^2=0$ (describing the future directed radial null geodesics)
one gets:
\begin{equation}
\Delta\chi=2\int_{\eta}^\delta\frac{\sin\eta}{\sin(\delta-\eta)}d\eta=+\infty.
\label{inf}
\end{equation}
We note that if $\delta=\pi$ then
$$
\Delta\chi=2(\pi-\eta)<\infty.
$$
The result (\ref{inf}) shows that radial null geodesics
circumnavigate the universe an infinite number of times as $t\to
t_f$. This fact and the homogeneity+isotropy results in conclusion
that (i) this universe has no event horizons and (ii) all future
endless timelike curves define the same c-boundary point. At last,
$$
\Delta\chi=2\int_0^{\eta}\frac{\sin\eta}{\sin(\delta-\eta)}d\eta<+\infty.
$$
One can show that all energy conditions are valid. For this let us point out
that sum $\rho+3p$~ in general model is equal to the sum $\rho+3p$~ in
starting model (see (\ref{inv}) for the case $n=1$). This fact results in
validity of strong energy condition for our model. Finally, from
Friedmann equations one can see that density of energy is always
positive at $k=1$. By this property and the validity of the strong energy
condition, the weak energy one will be satisfied automatically.

\subsection{\label{sec:level3} Generalization of a Lambda-radiation model in flat space}

Let's consider the flat universe which has a positive vacuum energy
$\Lambda$. Let's also assume that the universe is filled with the radiation.
Solving system (\ref{Fried}) we will obtain the initial solution for the scale
factor
\begin{equation}
\begin{array}{l}
    a=a_{0}\sinh^{1/2}\theta,\qquad
    \displaystyle{\theta=2\sqrt{\Lambda}t},\\ \\
\displaystyle{\rho=\Lambda(1+\sinh^{-2}\theta)},
\end{array}
\label{sol-2}
\end{equation}
where $a_{0}$ is a positive constant. If
$t>t_{v}=\sqrt{\frac{1}{4\Lambda}}{\rm arcsinh}(1)$, the strong
energy condition will necessarily be violated.  Using (\ref{an})
for the case $n=1$ one can see that general solution can be
written in the following form
\begin{equation}\label{GenSolLambda}
    a_{gen}\equiv a=a_0\sinh^{1/2}\theta
    \ln\frac{\coth^{\epsilon}\frac{\theta}{2}}{\delta},
\end{equation}
where $\delta$ is a positive constant. The parameter $\epsilon$, introduced here,
plays an important role in our reasonings. If $\epsilon=0$, then (\ref{GenSolLambda}) will
be equivalent to the initial solution. In the remaining cases one can
without any loss of generality assume $\epsilon=1$. There will be
three types of solutions. If $\delta<1$ the universe will be
open. This type of solutions has the following asymptotic behavior
\[a\rightarrow-0.5a_{0}\ln\delta\exp(\sqrt{\Lambda}t),\qquad
t\rightarrow\infty.\] It is easy to see that this universe is plagued by the
events horizon.

The case $\delta\geq 1$ is a more interesting one. If $\delta<<1$,
then solution will describe the universe, starting from an initial
singularity ($\theta=0$, $t=0$) and ending up in the final
singularity at $t_{f}=\sqrt{\frac{1}{\Lambda}}{\rm
arccoth}\delta$. One shall note that, if $\delta\geq 1+\sqrt{2}$
the strong energy condition will always be satisfied (in fact,
universe will end up in singularity long before the time $t_{v}$).

When $\delta=1$ the resulting solution might be denoted as
"quasisingular" because
\[\lim_{t\rightarrow \infty}a\sim
\lim_{\theta\rightarrow
\infty}\sinh^{1/2}\theta\ln\coth\frac{\theta}{2}\sim\exp(-\theta/2).\]
From this relation one can see that scale factor tends to
singularity but never achieves it.

Both singular and quasisingular models contains no events
horizons. To show this for the quasisingular case let us integrate
the equation for future directed radial null geodesics ($ds^2=0$) just like
it has been done in the previous subsection:
\[\Delta\chi\sim\int\limits_{0}^{\infty}\frac{d\theta}{\sinh^{1/2}\theta\ln\coth\frac{\theta}{2}}.\]
This integral diverges because subintegral expression has
an exponential asymptot at large $\theta$. Therefore radial null
geodesics circumnavigate the universe an infinite number of times
as $t\to \infty$. This fact and the homogeneity+isotropy result in
the conclusion similar to the one from the subsection B, namely that such universe
possess no events horizon. Absence of events horizon for singularity
model can be proved by analogy.

In conclusion let us analyze the equation of state for the generalization
of a lambda-radiation model. One can show that the value
$w=p/ \rho$ is equal to
\begin{equation}
w=-\frac{1}{3}+\frac{2}{3}\frac{(1-\sinh^{2}\theta)
\ln^{2}\frac{\coth\frac{\theta}{2}}{\delta}}{(\cosh\theta\ln\frac{\coth\frac{\theta}{2}}{\delta}-2)^{2}}.
\label{W}
\end{equation}
From this relation it follows immediately that for both open and quasisingular
models $w\rightarrow-1$ at large $t$ (large $\theta$). For the closed
model $w=-1/3$ whenever we approach the final singularity. In the initial
singularity $w=1/3$ for all models. Close examination of equation (\ref{W})
shows that $w$ is always greater than -1 for all cases, i.e. weak
energy condition will always be satisfied.

\section{Possible generalizations of the method}
In the previous sections we have shown that no finite-time observations made by our astronomer will yield an exact picture of the future dynamics of her universe. One can claim, however, that the cited "nonuniqueness" is due to the special properties of the Friedmann equations and will probably disappear for the cosmological models, distinct from this one. In current section we are going to show that, in fact, the above-mentioned ideology can be applied absolutely similarly in at least two particular cases: for anisotropic universes and for a certain brane world models.
\subsection{\label{sec:leve21} Anisotropic universes}
In this subsection we are going to show that the described method can be generalized to some cases of
anisotropic universes. Consider a simplest possible case of anisotropic model:
$$
ds^2=dt^2-a^2(t)dx_1^2- b^2(t)dx_2^2- c^2(t)dx_3^2,
$$
with the perfect fluid energy-momentum tensor. The corresponding system of equation has the form
\begin{equation}
\begin{array}{l}
\displaystyle{
\frac{\ddot a}{a}+\frac{\dot a}{a}\left(\frac{\dot b}{b}+\frac{\dot c}{c}\right)=\frac{3}{2}
\left(\rho-p\right),}\\
\\
\displaystyle{
\frac{\ddot b}{b}+\frac{\dot b}{b}\left(\frac{\dot a}{a}+\frac{\dot c}{c}\right)=\frac{3}{2}\left(\rho-p\right),}\\
\\
\displaystyle{ \frac{\ddot c}{c}+\frac{\dot c}{c}\left(\frac{\dot
a}{a}+\frac{\dot b}{b}\right)=\frac{3}{2}\left(\rho-p\right),}
\end{array}
\label{aniz-1}
\end{equation}
and
\begin{equation}
\displaystyle{ \frac{\ddot a}{a}+\frac{\ddot b}{b}+\frac{\ddot
c}{c}=-\frac{3}{2}\left(\rho+3p\right)}. \label{aniz-2}
\end{equation}
Using (\ref{aniz-1}) one can rewrite (\ref{aniz-2}) as
\begin{equation}
{\dot a}{\dot b}c+{\dot a}{\dot c}b+a{\dot b}{\dot c}=3\rho abc.
\label{aniz-2-1}
\end{equation}
There are known quite a few exact solutions of the system (\ref{aniz-1}),
(\ref{aniz-2}), such as the vacuum solution of Kasner \cite{Kasner}, or dust solution
of Saf\"ucking, Heckmann  \cite{Schucking}. Our aim, however, is to develop the dressing procedure
for the (\ref{aniz-1}), (\ref{aniz-2}).

In order to do this lets first define a new function
$\psi=abc$. Using (\ref{aniz-1}) one can check this function
to be a solution of the Schr\"odinger equation:
\begin{equation}
{\ddot \psi}=V\psi, \label{aniz-3}
\end{equation}
with potential $V=9(\rho-p)/2$. In fact, this is just a replica of the case
$n=3$ of the Friedmann equation.

The general solution of the equation (\ref{aniz-3}) is
\begin{equation}
\Psi=c_1\psi+c_2{\hat\psi}, \label{aniz-4}
\end{equation}
where ${\hat\psi}$ is linearly independent counterpart of  $\psi$.
Since $\Psi$ is the solution of the (\ref{aniz-3}) with initial
potential $V$ (with the new density and pressure $\tilde\rho$  and
$\tilde p$, but same difference: $ {\tilde\rho}-{\tilde
p}=\rho-p$), we can define the new solutions of the system
(\ref{aniz-1}) $A$, $B$, $C$ with similar right hand side:
\begin{equation}
\displaystyle{ \frac{\ddot A}{A}+\frac{\dot A}{A}\left(\frac{\dot
B}{B}+\frac{\dot C}{C}\right)=\frac{3}{2}\left(\rho-p\right),}
\label{aniz-1A}
\end{equation}
\begin{equation}
\displaystyle{ \frac{\ddot B}{B}+\frac{\dot B}{B}\left(\frac{\dot
A}{A}+\frac{\dot C}{C}\right)=\frac{3}{2}\left(\rho-p\right),}
\label{aniz-1B}
\end{equation}
\begin{equation}
\displaystyle{ \frac{\ddot C}{C}+\frac{\dot C}{C}\left(\frac{\dot
A}{A}+\frac{\dot B}{B}\right)=\frac{3}{2}\left(\rho-p\right),}
\label{aniz-1C}
\end{equation}
and
\begin{equation}
{\dot A}{\dot B}C+{\dot A}{\dot C}B+A{\dot B}{\dot
C}=3{\tilde\rho}\Psi. \label{aniz-2-2}
\end{equation}
Further on, one can represent equations (\ref{aniz-1A}),
(\ref{aniz-1B}) as the particular system
\begin{equation}
\begin{array}{l}
2A{\ddot A}\Psi+2A{\dot A}{\dot \Psi}-2\Psi{\dot A}^2+3(p-\rho)A^2\Psi=0,\\
2B{\ddot B}\Psi+2B{\dot B}{\dot \Psi}-2\Psi{\dot
B}^2+3(p-\rho)B^2\Psi=0.
\end{array}
\label{aniz-5}
\end{equation}
One can show that if $A$, $B$ and $\Psi$ are solutions of respectively
(\ref{aniz-5}) and (\ref{aniz-3}) then the function
\begin{equation}
C=\frac{\Psi}{AB}, \label{aniz-C}
\end{equation}
would be a solution of (\ref{aniz-1C}). Thus, assuming $\rho$, $p$ and $\Psi$
to have certain values, and solving (\ref{aniz-5}) for
$A$ and $B$, one can automatically calculate $C$ (via (\ref{aniz-C})). The
new density $\tilde\rho$ will then be defined from the eq. (\ref{aniz-2-2})
and the new pressure from
$$
{\tilde p}={\tilde\rho}-\rho+p.
$$
Equations (\ref{aniz-5}) can be solved by simple
substitution $A=\exp(\alpha)$, $B=\exp(\beta)$ which results in
system
$$
\begin{array}{l}
\displaystyle{
{\ddot \alpha}+\frac{{\dot\Psi}}{\Psi}{\dot\alpha}=\frac{3}{2}(\rho-p)},\\
\\
\displaystyle{ {\ddot
\beta}+\frac{{\dot\Psi}}{\Psi}{\dot\beta}=\frac{3}{2}(\rho-p)},
\end{array}
$$
therefore
$$
A=F(t)\exp\left(c_{_I}\int^t\frac{dt'}{\Psi(t')}\right),
$$
and $$
 B=F(t)\exp\left(c_{_{II}}\int^t\frac{dt'}{\Psi(t')}\right),
$$
where
$$
F(t)=\exp\left[\frac{3}{2}\int^t\frac{dt'}{\Psi(t')}\int^{t'}dt''\left(\rho(t'')-p(t'')\right)\right],
$$
$c_{_I}$ and $c_{_{II}}$ are arbitrary constants.

\subsection{\label{sec:leve22} A Few notes about the brane universe}

Let's start our considerations from the following system:
\begin{equation}
\begin{array}{cc}
\displaystyle{\left(\frac{\dot{a}}{a}\right)^2=\rho
\left(1+\frac{\rho}{2 \lambda}\right),}
\\
\\
\displaystyle{-2\frac{\ddot{a}}{a}=\rho+3p+\frac{\rho}{\lambda}\left(2
\rho+3p \right),}
\\
\\
\end{array}
\label{novobran}
\end{equation}
where $\rho$ is the density, $p$ - pressure, and $\lambda$ is the
tension on the brane.

Before we actually start applying our method, let us make one
simple but interesting observation. If we'll try to solve the
system (\ref{novobran}) for the $\rho$ as a function of $a$ (for
finite values of $\lambda \neq 0$), we will unexpectedly end up
with not just one, but \textit{two} distinct solutions:
\begin{equation}
\rho_{1,2}=-\lambda \pm \sqrt{\lambda^2+2 \lambda
\left(\frac{\dot{a}}{a}\right)^2}
\end{equation}

For the negative values of $\lambda$ this (interestingly enough)
gives an upper bound on the absolute value of the Hubble constant:
$H^2=\left(\frac{\dot{a}}{a}\right)^2 \leq -\lambda/2$ and
presents two positive values of $\rho$. In the case of positive
$\lambda$ (which in theory might correspond to our universe) one
of the values of $\rho$ will be negative (it might seem strange on
a first glance, but the study of the models with the negative
$\rho$ has experienced quite a surge in a past few years. See, for
example \cite{Bigs}). In other words, even the exact knowledge of
the very functional representation of scale factor would not allow
the astronomer (living in such universe) to uniquely determine the
density function!

With this being said let us finally state the following version of our linearization theorem:

{\bf Linearization Theorem for Branes}. Let $a=a(t)$ be the solution of (\ref{novobran}) with
$k=0$ and with $\rho$ and $p$~ being the density and pressure functions correspondingly. Then

1. Function $\psi_n\equiv a^n$ will be solution of the Schr\"odinger equation
\begin{equation}
\frac{{\ddot \psi_n}}{\psi_n}=W_n, \label{Scrone}
\end{equation}
with the potential
\begin{equation}
\begin{array}{l}
\displaystyle{ W_n=\frac{n}{2}\left(2n \rho-3\left(\rho+p
\right)+\frac{\rho}{\lambda}\left(n
\rho-3\left(\rho+p\right)\right)\right)=}\\
\\
\displaystyle{=\frac{n}{2}\left[2n(K+V)-6K+\frac{1}{\lambda}(K+V)\left(n(K+V)-6K\right)\right]};
\end{array}
 \label{Wn}
\end{equation}
(Note, that despite the discussed non-uniqueness of the density
function, potential $W_n$ is unique!)

2. The two-parameter function $a_n=a_n(t;c_1,c_2)$, satisfying (\ref{an}) will be solution of (\ref{novobran}) with new energy density $\rho_n$ and pressure $p_n$ satisfying:
\begin{equation}
W_n(\rho,p)=W_n(\rho_n,p_n).
\end{equation}
As before, the proof is nothing more but a technical exercise that will be left out for our readers.
\newline
\newline
{\bf  Example}. System (\ref{novobran}) can easily be solved when $p/\rho=w=$const. Let  $\lambda>0$, $w>-1$ ($\gamma>0$)
and $a(0)=0$ then
\begin{equation}
\begin{array}{l}
\displaystyle{ a(t)=a(t_0)\left(\frac{t(3\lambda\gamma
t+2\sqrt{2\lambda})}{t_0(3\lambda\gamma
t_0+2\sqrt{2\lambda})}\right)^{1/3\gamma}},\\
\\
\displaystyle{H=\frac{\dot a}{a}=\frac{2(3\sqrt{\lambda}\gamma
t+\sqrt{2})}{3\gamma t(3\sqrt{\lambda}\gamma t+2\sqrt{2})},}\\
\\
\displaystyle{ \rho=\frac{4\lambda}{3\gamma
t(3\sqrt{\lambda}\gamma t+2\sqrt{2})}}.
\end{array}
\label{brane-solution}
\end{equation}
For the dust case $w=0$ ($\gamma=1$) and $n=3$ one gets the following
two-parameter function $a_3=a_3(t;c_1,c_2)$:
\begin{equation}
a_3=\left[\left(c_1-\frac{3c_2\sqrt{2}}{\sqrt{\lambda}}
\log\left(\frac{\zeta(t)}{t}\right)\right)t\zeta(t)+2c_2\left(\frac{\zeta(t)}
{\lambda}+3t\right)
\right]^{1/3}, \label{ogogo-1}
\end{equation}
where $\zeta(t)=3\lambda t+2\sqrt{2\lambda}$. One can show that in
this case the solution might above all describe the space-times without
events horizon (for a special choice of values of $c_1$, $c_2$).
\newline
{\it Remark 5}. If $\gamma<0$ then one get the phantom model with
two big rips \cite{Bigs}:
$$
a(t)=\left(\frac{\mu^2}{1-4.5\lambda\gamma^2(t-T)^2}\right)^{1/3|\gamma|},
$$
where $\mu^2$ and $T$ are constants. Using our method one can
obtain two-parameter (or $2N$-parameter) function
$a_n=a_n(t;c_1,c_2)$ such that $a_n(t;c_1=1,c_2=0)=a(t)$.

\section{Conclusion}

In this paper we have discussed a simple (and easily automatizable) method of construction of exact solutions of Friedmann equations. Despite simplicity, the method allows for acquirement of solutions characterized by extremely interesting properties. What is more, it appears (for $k$=0) that the very abundance of the set of solutions that are to be obtained this way leads us to a stunning conclusion: no matter how accurate our astronomical observations are, there exist not just one, but a whole set of solutions that will satisfy the observational data while leading to essentially different dynamics in future. This sudden twist leads us to seemingly unavoidable conclusion about the principle indefiniteness of the future, hidden in the Fridmann equations. For a first glance such conclusion looks really disappointing, rendering useless
all our efforts to build a suitable cosmological model describing our universe.

However, everything above-said doesn't mean the impossibility to
determine the actual dynamics of the universe \textbf{in principle}. Even though the
usual observational methods might not give us the final answer, we can use
the statistical methods to get it - that is, anthropic principle.

For example, in the article we have studied the particular models
possessing the final singularity and no event horizons. What is
the practical significance of such models? As follows from the
recent observations (see  \cite{Riess}, \cite{Perlm}) our universe
suffers the accelerated expansion \cite{IJMPD}, \cite{Chern}. As
for now, the most probable cause of such expansion lies in nonzero
cosmological constant. If this is really the case, then the future
dynamics of observable universe is confined within the particles
horizon and, as such, leads to problems with formulation of a
fundamental physical theory (like the string theory or
hypothetical M-theory) in a finite volume \cite{Banks}. If, on the
other hand, the lifetime of the universe (with an observable
expansion rate) will exceed the limit of $10^{60}$ years, the
dominating observers will, as follows from \cite{Don}, be the ones
of a quantum fluctuations origin, which, of course, could hardly
be called compatible with our observations. Therefore, as follows
from the anthropic reasons of \cite{Don}, the most probable
scenarios would be those describing the contemporary expansion,
being traced by the consequent contraction phase and the
``horizonless'' collapse~\footnote{Another way to avoid this
conclusion is to suppose that our universe has a certain decay
rate for tunneling into oblivion: Don N. Page, [hep-th/0612137];
Don N. Page, [hep-th/0610079].}. In other words, exactly that type
of scenarios, that has been naturally constructed with the aid of
``dressed'' initial solutions.

Finally, let us emphasize that the purpose of this paper was a mere study of Friedmann equation's solutions that has been constructed in this article by the linearization procedure. The practical aspects of application of the described technique to the observed universe lied out of scoop of this work. However, we are hoping to get back to it (as well, as to applications of the method to more interesting anisotropic and brane models) in one of our future articles.

\acknowledgements

\noindent  We would like to thank  A. A. Starobinsky for important
notices he made on the first draft of this paper. The idea of generalization of our method into the
anisotropic cosmological models has been proposed (in the private discussion) by J.D.Barrow, to whom we give our utmost gratitudes. Finally, authors have certainly benefited from discussions with A. A. Yurova whom we would like to
thank too.

\end{document}